\documentclass[11pt,english]{article}
\usepackage[T1]{fontenc}
\usepackage[latin1]{inputenc}
\usepackage{a4wide}
\usepackage{amsmath}
\usepackage{color}
\usepackage{graphicx}
\usepackage{amssymb}

\makeatletter

\providecommand{\tabularnewline}{\\}

\usepackage{babel}
\makeatother
\begin{document}
$\,\,$
\vspace{1cm}

\begin{center}\textbf{\LARGE BPS $Z_{N}$ String Tensions, Sine law
and Casimir}\end{center}{\LARGE \par}

\begin{center}\textbf{\LARGE Scaling and Integrable Field Theories}\end{center}{\LARGE \par}

\begin{center}

\vspace{1cm}

{{\bf Marco A. C. Kneipp}} 

\vspace{0.3cm}

\end{center}

\begin{center}{\em Universidade Federal de Santa Catarina (UFSC)%
\footnote{Permanent address. E-mail: kneipp@fsc.ufsc.br.%
},\\ 

Departamento de F\'\i sica, CFM,\\

Campus Universit\'ario, Trindade,\\

88040-900, Florian\'opols, Brazil.\\ 
\medskip{}

International Centre for Theoretical Physics (ICTP), \\ 

Strada Costiera 11, \\

34014, Trieste, Italy. \\ }

\end{center}

\begin{abstract}
We consider a Yang-Mills-Higgs theory with spontaneous symmetry breaking
of the gauge group $G\rightarrow U(1)^{r}\rightarrow C_{G}$, with
$C_{G}$ being the center of $G$. We study two vacua solutions of
the theory which produce this symmetry breaking. We show that for
one of these vacua, the theory in the Coulomb phase has the mass spectrum
of particles and monopoles which is exactly the same as the mass spectrum
of particles and solitons of two dimensional affine Toda field theory,
for suitable coupling constants. That result holds also for ${\cal N}=4$
Super Yang-Mills theories. On the other hand, in the Higgs phase,
we show that for each of the two vacua the ratio of the tensions of
the BPS $Z_{N}$ strings satisfy either the Casimir scaling or the
sine law scaling for $G=SU(N)$. These results are extended to other
gauge groups: for the Casimir scaling, the ratios of the tensions
are equal to the ratios of the quadratic Casimir constant of specific
representations; for the sine law scaling, the tensions are proportional
to the components of the left Perron-Frobenius eigenvector of Cartan
matrix $K_{ij}$ and the ratios of  tensions are equal to the ratios
of the soliton masses of affine Toda field theories. 

\vfill PACS numbers: 11.27.+d, 11.15.-q, 02.20.Sv 

\thispagestyle{empty}
\end{abstract}
\newpage

\section{Introduction}

In $SU(N)$ QCD, it is believed that the confinement of particles
in strong coupling regime happens by formation of chromoelectric flux
tubes, which we shall call QCD strings, carrying charge on the discrete
group $Z_{N}$. There are different ways to try to understand this
phenomenon (for nice reviews see \cite{strassler}\cite{greensite}).
In particular it is believed that particle confinement in strong coupling
could be a phenomenon dual to the monopole confinement in weak coupling.
For some time, it was thought that QCD string could be dual to strings
solutions appearing in (effective) theories with broken $U(1)$ gauge
group. However it seems it does not give the right spectrum of mesons\cite{strassler}.
For this reason, we have studied many properties of topological $Z_{N}$
strings solutions and monopole confinement in the Higgs phase of theories
with simple gauge groups $G$ (\textit{without} $U(1)$ factors) \cite{kbrockill2001}\cite{k2002}\cite{k2003}.
More recently some works \cite{G x U(1)} also appeared analyzing
semi-local strings\cite{vachaspati} with gauge group $SU(N)\times U(1)$
and global flavor symmetry $SU(N)_{\textrm{flavor}}$ .

Our motivation for considering chromomagnetic $Z_{N}$ strings in
the Higgs phase produced by breaking of gauge groups $G$ without
$U(1)$ factors is that the QCD's chromoelectric strings in confining
phase should be formed only by fields with $SU(3)$ color charges
and not $U(1)$ gauge fields. We consider general gauge groups $G$
since it allows us to consider more general and direct arguments which
may clarify some fundamental results common to different groups. We
also hope that these results might be useful for lattice calculation
of chromoelectric strings for groups other than $SU(N)$. 

Differently  from the non-Abelian semi-local string solutions with
gauge groups with $SU(N)\times U(1)$ where the tension is only due
magnetic flux in the $U(1)$ direction \cite{G x U(1)} and it depends
on the $U(1)$ winding number, for the $Z_{N}$ strings obtained by
breaking simple gauge groups $G$, we showed that they are associated
to weights of representations of the dual group $G^{\vee}$ and the
string tensions seems to be consistent with the tensions of QCD strings
as discussed in \cite{k2003} and in the present work. Therefore,
the chomomagnetic $Z_{N}$ string solutions we consider, have features
similar to QCD strings. In particular in \cite{k2003}, there were
constructed $Z_{N}$ strings solutions which appear in a theory with
symmetry breaking pattern\begin{equation}
G\,\stackrel{\phi_{1}^{{\scriptstyle \textrm{vac}}}}{\rightarrow}\, U(1)^{r}\,\stackrel{\phi_{2}^{{\scriptstyle \textrm{vac}}}}{\rightarrow}C_{G},\label{1}\end{equation}
 where $r$ is the rank of $G$ and $C_{G}$ its center. For each
weight of $G$ one can construct a $Z_{N}$ string and we established
how these strings are separated into topological sectors for general
$G$. The string flux quantization condition was also obtained and
the flux matching between strings and monopoles for \textit{any} group
$G$ was shown. The \textit{set} of strings which should be attached
to each monopole was shown to belong necessarily to the trivial topological
sector which is consistent with the fact that only these configurations
can terminate at some point and can break as it happens for the QCD
strings. In particular, for $G=SU(N)$, a string solution was constructed
for each weight (color) of the $N$ dimensional fundamental representation.
We determined the set of strings which should be attached to each
non-Abelian monopole which appeared in the Coulomb phase and showed
that one could also have a confining system composed by $N$ monopoles
besides the monopole-antimonopole system. Since these monopoles have
magnetic charges in the adjoint representation of the dual group $G^{\vee}$\cite{MontoneOlive},
they should be dual to gluons and these confined systems should be
dual to the glueballs. Differently from what is sometimes said, the
$Z_{N}$ strings does not necessarily point in a direction in the
Cartan subalgebra (CSA). However, since the monopoles´ magnetic flux
is in the direction of the CSA \cite{gno}, we only consider $Z_{N}$
string solutions with flux in the CSA which are the relevant for confinement
of these monopoles. This result is analogous to the Abelian dominance
observed in the confined phase of QCD.

An important quantity in particle confinement in QCD are the string
tensions. The spectrum of string tensions has been extensively studied
in lattice calculations in recent years \cite{lattice}\cite{latticeteper}.
The main conjectures for the QCD string tensions are the Casimir scaling\cite{casimir}
and the sine law scaling\cite{douglas shenker}. In supersymmetric
theories it was possible to arrive to these scalings by using analytical
calculation. For softly broken ${\cal N}=2$ super Yang-Mills theories
with gauge group $G=SU(N)$ and a hypermultiplet in fundamental representation
it was obtained an effective Lagrangian with spontaneouly broken $U(1)^{N-1}$
Abelian gauge group and Nielsen-Olesen strings\cite{NielsenOlesen}
with tensions satisfying the sine law scaling \cite{douglas shenker}.
On the other hand, in \cite{k2003} a softly broken ${\cal N}=4$
Super Yang-Mills theory was considered and obtained the Casimir scaling
 for tensions of the BPS $Z_{N}$ strings when $G=SU(N)$. The sine
law scaling was also derived in the M theory description of ${\cal N}=1$
$SU(N)$ Super Yang-Mills theory\cite{HananyStrasslerZaffaroni} and
in the AdS/CFT correspondence \cite{HerzogKlebanov}. 

In the present work we show that for the BPS $Z_{N}$ strings, one
can obtain the Casimir scaling and the sine law scaling by considering
two different vacua of the same theory which give rise to the symmetry
breaking (\ref{1}). This result shows that these scalings are not
necessarily {}``universal laws'', but they depend on the vacuum
which is responsible for the symmetry breaking. From the dual superconductor
picture this result may indicate that \textcolor{black}{if the tensions
of the QCD strings satisfy one of these scalings, it may be due to
a non-Abelian monopole condensate (in the adjoint representation)
in one of these two vacua.} We also generalize the Casimir and sine
law scalings to groups other than $SU(N)$. It is important to note
that in \cite{douglas shenker}, the sine law scaling was obtained
for the tensions of $Z$ strings which appear due to the spontaneouly
broken $U(1)^{N-1}$ Abelian gauge group, which gives rise to a different
meson spectrum \cite{strassler} from $Z_{N}$ strings we consider.
Since in this paper we are interested in studying some general properties
at the classical level of these $Z_{N}$ strings which may be useful
for QCD and not necessarily confinement in supersymmetric theories,
we shall not restrict the potential to be supersymmetric. Similarly
to the monopole solutions, for the $Z_{N}$ string one can construct
moduli spaces of solutions. However, for the determination of the
properties of the $Z_{N}$ string solution as string flux and tension,
it is not necessary to construct moduli spaces since for all solutions
in a moduli space these properties are the same. 

In this paper we introduce, in sections 2 and 3 some general results
for BPS $Z_{N}$ string and Lie algebra which will be used in the
following sections. In section 4 we obtain two different vacuum solutions
which give rise to the symmetry breaking (\ref{1}) for any group
$G$. The first stage of the symmetry breaking corresponds to the
Coulomb phase which is analyzed in section 5. In particular we show
that for one of the vacua, the mass spectrum of particles and monopoles
of the four dimensional theories in the Coulomb phase is exactly the
same as the mass spectrum of particles and solitons of two dimension
affine Toda field theories (ATFTs), if the couplings of the two theories
satisfy some suitable relations. That result holds also for ${\cal N}=4$
Super Yang-Mills theories. Some other relations between BPS $Z_{N}$
strings and Affine Toda field theories are discussed in \cite{integrability Z(N) strings}.
Then, in section 6 we analyze the Higgs phase. We start reviewing
the construction of the $Z_{N}$ strings solutions, where for each
weight of the dual gauge group $G^{\vee}$ one can construct a solution,
and how these solutions are classified in topological sectors for
any group $G$. Then, we obtain the BPS string tension for each topological
sector and show that, depending on the vacuum, the ratios of the tensions
satisfy the sine law scaling or the Casimir scaling when the gauge
group is $SU(N)$. These scalings are generalized to other groups,
and in particular the tensions which appear in the sine law scaling
are identified with components $x_{i}^{(1)}$ of left Perron-Frobenius
eigenvector of the Cartan matrix $K_{ij}$ and the ratios of tensions
are equal to the ratios of soliton masses of the correponding affine
Toda field theory, for any gauge group $G$. Differently from the
$SU(n)$ group which the Casimir and the sine law scaling coincide,
at fixed $k$ and large $n$, for $G=Spin(2n)$ (the universal covering
group of $SO(2n)$), the Casimir and the sine law scaling give different
results in the large $n$ limit. With the generalization of the Casimir
and sine law scaling for the $Z_{N}$ strings to other gauge groups,
it could be interesting to analyze, using lattice calculation, the
chromoelectric string tensions of QCD for groups other than $SU(n)$,
as for example $G=Spin(2n)$. It is important to note that the Casimir
scaling and the sine law scaling we obtained are lower bounds for
the non-BPS $Z_{N}$ string tensions and they hold exactly only for
the BPS $Z_{N}$ strings, which exist only on the boundary between
a type I and type II superconductor. Therefore, the small deviation
from the Casimir scaling observed in \cite{latticeteper} could be
due to fact that QCD strings would not be BPS. Recently the interest
on topological solutions in (supersymmetric) field theory \cite{solitons}
has increased. We hope that the results may also be useful for the
study of other topological solutions

\section{BPS $Z_{N}$ strings}

Let us consider Yang-Mills-Higgs theories with arbitrary gauge group
$G$ which is simple, connected and simply connected. In order to
exist strings and confined monopoles we shall consider theories with
two complex scalars fields%
\footnote{Note that, if one only wants just string solutions, it is enough only
one complex scalar.%
} $\phi_{s},\, s=1,2$, in the adjoint representation of $G$. We also
consider that the vacuum solutions $\phi_{1}^{\textrm{vac}}$, $\phi_{2}^{\textrm{vac}}$
produce the symmetry breaking (\ref{1}). 

In order to exist stable $Z_{N}$ string solutions for the symmetry
breaking (\ref{1}), $C_{G}$ must be non-trivial. Therefore we shall
not consider the groups $E_{8}$, $F_{4}$ and $G_{2}$. In table
1 we list the centers of simply connected simple Lie groups. In \cite{kbrockill2001}\cite{k2002}
we consider an alternative symmetry breaking in which stable strings
exist even in theories with gauge groups with trivial center. 

\begin{table}
\begin{center}\begin{tabular}{|c|c|c|c|}
\hline 
G&
Extended Dynkin diagram of $g$&
$W_{0}$&
$C_{G}$\tabularnewline
\hline
\hline 
$SU(n+1)$&
\includegraphics[%
  bb=0bp 0bp 331bp 128bp,
  scale=0.3]{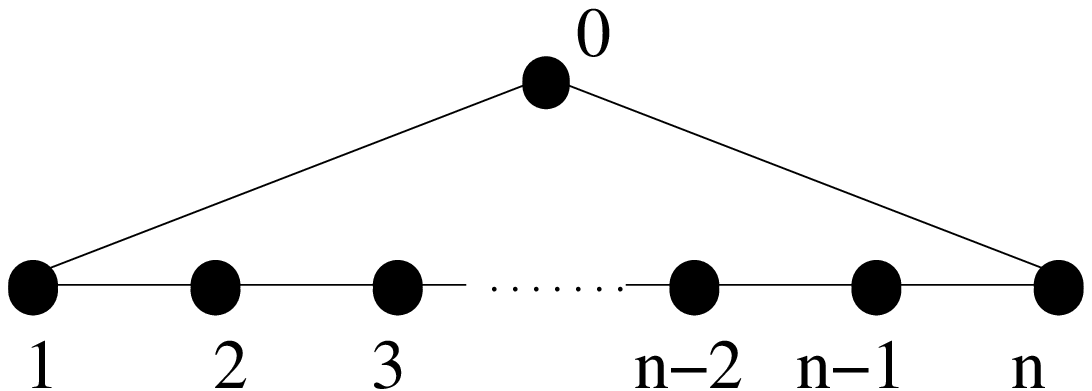}&
$0,1,2,...,n$&
$\mathbb{Z}_{n+1}$\tabularnewline
\hline 
$Spin(2n+1)$&
\includegraphics[%
  bb=0bp 0bp 305bp 128bp,
  scale=0.3]{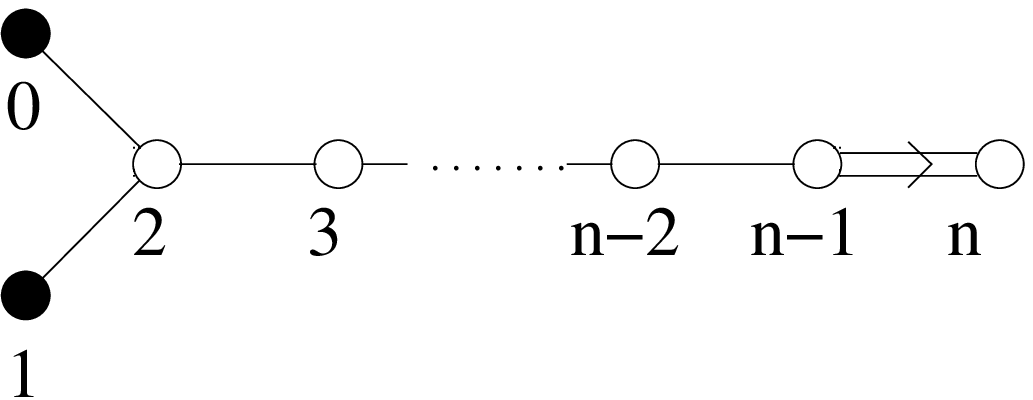}&
$0,1$&
$\mathbb{Z}_{2}$\tabularnewline
\hline 
$Sp(2n)$&
\includegraphics[%
  bb=0bp 0bp 319bp 67bp,
  scale=0.3]{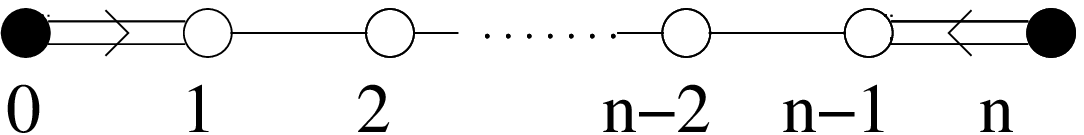}&
$0,n$&
$\mathbb{Z}_{2}$\tabularnewline
\hline 
$Spin(4n)$&
\includegraphics[%
  bb=0bp 0bp 328bp 129bp,
  scale=0.3]{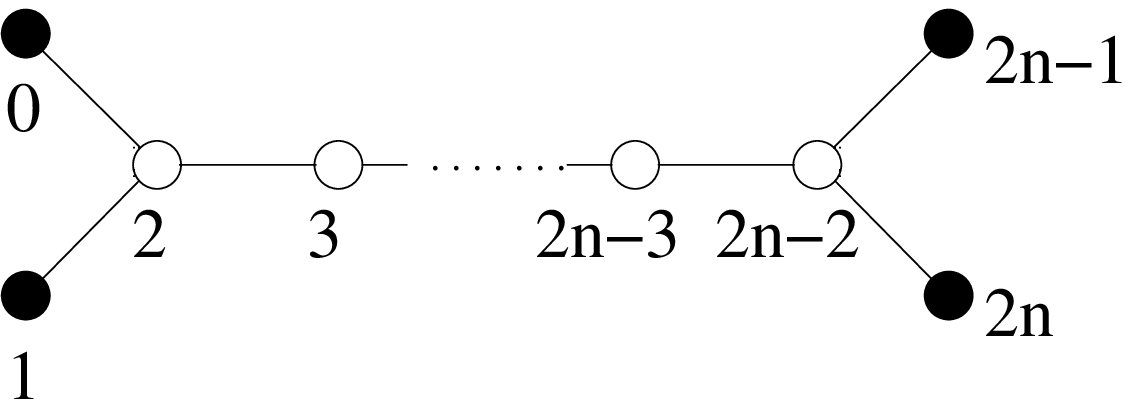}&
$0,1,2n-1,2n$&
$\mathbb{Z}_{2}\times\mathbb{Z}_{2}$\tabularnewline
\hline
$Spin(4n+2)$&
\includegraphics[%
  bb=0bp 0bp 332bp 133bp,
  scale=0.3]{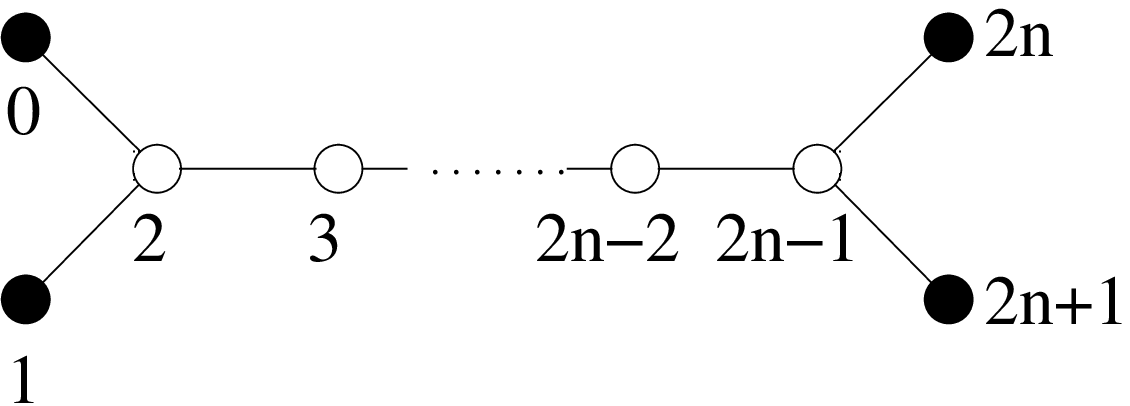}&
$0,1,2n,2n+1$&
$\mathbb{Z}_{4}$\tabularnewline
\hline
$E_{6}$&
\includegraphics[%
  bb=0bp 0bp 232bp 160bp,
  scale=0.3]{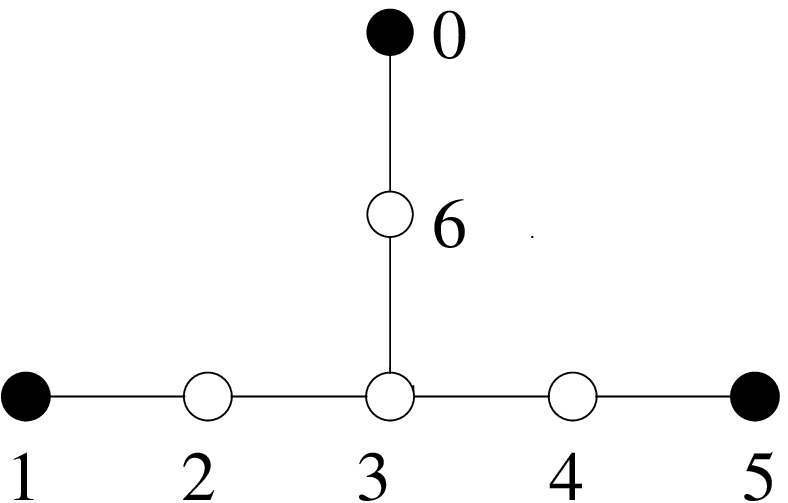}&
$0,1,5$&
$\mathbb{Z}_{3}$\tabularnewline
\hline 
$E_{7}$&
\includegraphics[%
  bb=0bp 0bp 394bp 105bp,
  scale=0.3]{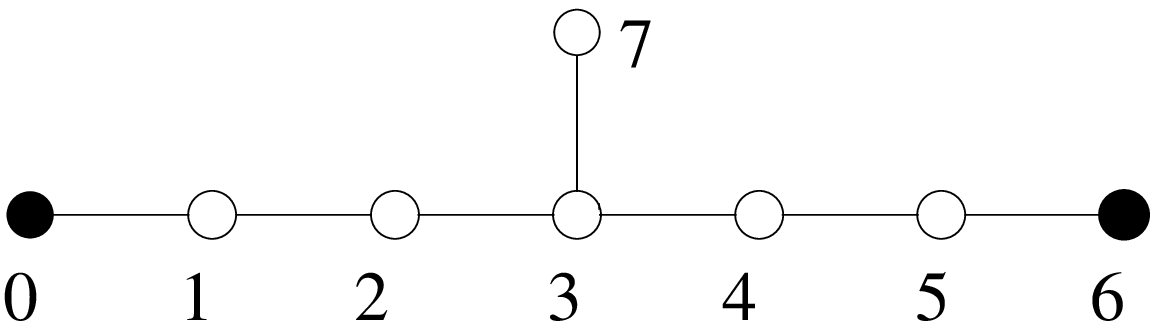}&
$0,6$&
$\mathbb{Z}_{2}$\tabularnewline
\hline
$E_{8}$&
\includegraphics[%
  bb=0bp 0bp 394bp 105bp,
  scale=0.3]{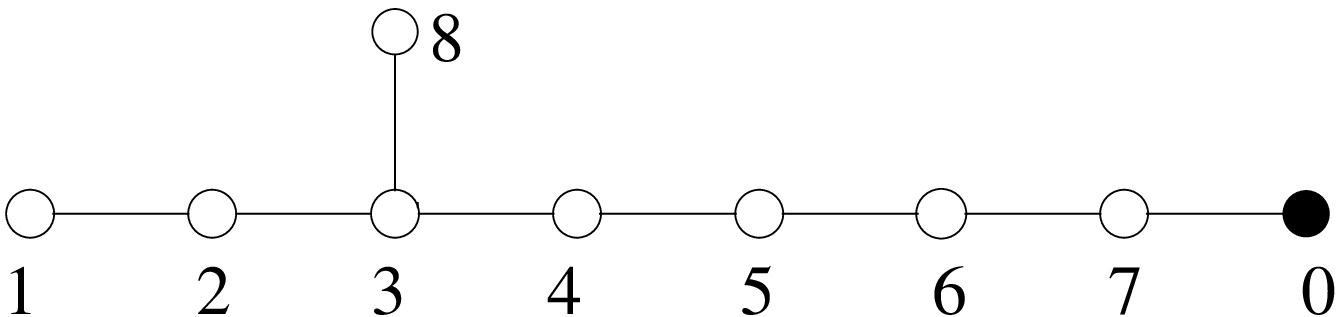}&
$0$&
$1$\tabularnewline
\hline 
$F_{4}$&
\includegraphics[%
  bb=0bp 0bp 232bp 66bp,
  scale=0.3]{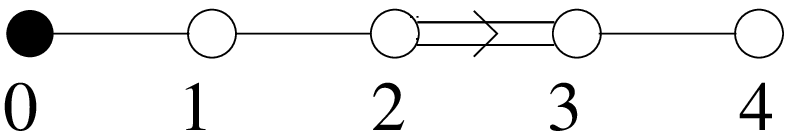}&
$0$&
$1$\tabularnewline
\hline 
$G_{2}$&
\includegraphics[%
  bb=0bp 0bp 122bp 70bp,
  scale=0.3]{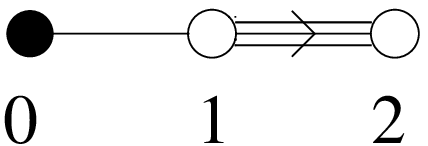}&
$0$&
$1$\tabularnewline
\hline
\end{tabular}\end{center}

\caption{\label{cap:Extended-Dynkin}\label{Dynkin}Extended Dynkin diagrams,
nodes symmetrically related to the node $0$ and center groups $C_{G}$. }
\end{table}

The Lagrangian of the theory we are to study is\begin{equation}
\mathcal{L}=-\frac{1}{4}G_{a\mu\nu}G_{a}^{\mu\nu}+\frac{1}{2}\left(D_{\mu}\phi_{s}\right)_{a}^{*}\left(D^{\mu}\phi_{s}\right)_{a}-V(\phi,\phi^{*})\label{4}\end{equation}
where $D_{\mu}=\partial_{\mu}+ie[W_{\mu},$~~{]}. Let $D_{\pm}=D_{1}\pm iD_{2}$
and $B_{ai}=-\epsilon_{ijk}G_{ajk}/2$ is the non-Abelian magnetic
field. As analyzed in \cite{kbrockill2001}\cite{k2003}, the BPS
string conditions for a theory with gauge group $G$ without $U(1)$
factors, are\begin{eqnarray}
B_{a3} & = & \mp d_{a},\label{5a}\\
D_{\mp}\phi_{s} & = & 0,\label{5b}\\
V(\phi,\phi^{*})-\frac{1}{2}\left(d_{a}\right)^{2} & = & 0,\label{5c}\\
E_{ai} & = & B_{a1}=B_{a2}=D_{0}\phi_{s}=D_{3}\phi_{s}=0,\label{5d}\end{eqnarray}
with \[
d_{a}=\frac{e}{2}\left(\phi_{sb}^{*}if_{abc}\phi_{sc}\right)-X_{a}\]
where $X_{a}$ is a real scalar quantity which transforms in the adjoint
representation with dimension of mass. From its transformation properties
we can consider \begin{equation}
X_{a}=\frac{em}{2}Re(\phi_{1a}),\label{5e}\end{equation}
 where $m$ is a mass parameter which is considered to be non negative.
This term was introduced in \cite{kbrockill2001} as a generalization
of the Fayet-Iliopolous term in the sense that it is responsible for
the symmetry breaking which gives rise to stable string solutions
for theories with non-Abelian gauge groups. In this case, the string
tension satisfies\cite{kbrockill2001}\cite{k2003}\begin{equation}
T\geq\frac{me}{2}\left|\phi_{1}^{\textrm{vac}}\right|\left|\Phi_{\textrm{st}}\right|\label{7a}\end{equation}
where \begin{equation}
\Phi_{\textrm{st}}=\frac{1}{\left|\phi_{1}^{\textrm{vac}}\right|}\int d^{2}x\left[Re\left(\phi_{1}\right)_{a}B_{3a}\right]\label{7b}\end{equation}
is the string flux, with the integral being taking in the plane orthogonal
to the string. The equality in Eq. (\ref{7a}) happens only for the
BPS strings satisfying the conditions (\ref{5a})-(\ref{5d}). In
order to fulfill (\ref{5c}), we shall consider \begin{equation}
V(\phi,\phi^{*})=\frac{1}{2}\left(d_{a}\right)^{2}.\label{8}\end{equation}
Note that condition (\ref{5c}) does not restrict the potential to
have this form. In \cite{kbrockill2001}\cite{k2002}\cite{k2003}
it was considered different potentials.

\section{Mathematical results}

Let us start given some conventions and useful mathematical results
which will be used later on. Let $\mathfrak{g}$ be the Lie algebra
associated to the group $G$. Let us adopt the Cartan-Weyl basis in
which\begin{eqnarray*}
\textrm{Tr}\left(H_{i}H_{j}\right) & = & \delta_{ij},\\
\textrm{Tr}\left(E_{\alpha}E_{\beta}\right) & = & \frac{2}{\alpha^{2}}\delta_{\alpha+\beta},\end{eqnarray*}
 where the trace is taking in the adjoint representation. The generators
$H_{i},\, i=1,\,2,\,...,\, r$, form a basis for the Cartan subalgebra
(CSA) $\mathfrak{h}$. In this basis, the commutation relations read
\begin{eqnarray}
\left[H_{i},E_{\alpha}\right] & = & \left(\alpha\right)^{i}E_{\alpha},\label{3.1}\\
\left[E_{\alpha},E_{-\alpha}\right] & = & \frac{2\alpha}{\alpha^{2}}\cdot H,\nonumber \end{eqnarray}
where $\alpha$ are roots and the upper index in $\left(\alpha\right)^{i}$
means the component $i$ of $\alpha$. Then, $\alpha_{i}$ and $\lambda_{i}$,
$i=1,2,...,$r , are respectively the simple roots and fundamental
weights of $\mathfrak{g}$ and \begin{equation}
\alpha_{i}^{\vee}=\frac{2\alpha_{i}}{\alpha_{i}^{2}},\,\,\,\,\,\lambda_{i}^{\vee}=\frac{2\lambda_{i}}{\alpha_{i}^{2}}\label{3.1a}\end{equation}
are the simple coroots and fundamental coweights, which satisfy the
relations\begin{equation}
\alpha_{i}\cdot\lambda_{j}^{\vee}=\alpha_{i}^{\vee}\cdot\lambda_{j}=\delta_{ij}.\label{3.1b}\end{equation}
$\alpha_{i}^{\vee}$ and $\lambda_{i}^{\vee}$ are simple roots and
fundamental weights of the dual algebra $\mathfrak{g}^{\vee}$. Moreover,
\begin{equation}
\alpha_{i}=K_{ij}\lambda_{j}\label{3.3}\end{equation}
where \[
K_{ij}=\frac{2\alpha_{i}\cdot\alpha_{j}}{\alpha_{j}^{2}}\]
is the Cartan matrix associated to $\mathfrak{g}$.

The fundamental weights form a basis for the weight lattice of $G$,\begin{equation}
\Lambda_{w}(G)=\left\{ \omega=\sum_{i=1}^{r}n_{i}\lambda_{i},\,\,\,\,\,\,\, n_{i}\in\mathbb{Z}\right\} .\label{3.3a}\end{equation}
This lattice includes as a subset, the root lattice of $G$, \begin{equation}
\Lambda_{r}(G)=\left\{ \beta=\sum_{i=1}^{r}n_{i}\alpha_{i},\,\,\,\,\,\,\, n_{i}\in\mathbb{Z}\right\} ,\label{3.3b}\end{equation}
which has the simple roots $\alpha_{i}$ as basis. Similarly, the
fundamental coweights $\lambda_{i}^{\vee}$ are basis of the weight
lattice of the dual group%
\footnote{We shall consider the dual group $G^{\vee}$ as the covering group
associated to the dual algebra $\mathfrak{g}^{\vee}$.%
} $G^{\vee}$\begin{equation}
\Lambda_{w}(G^{\vee})=\left\{ \omega=\sum_{i=1}^{r}n_{i}\lambda_{i}^{\vee},\,\,\,\,\,\,\, n_{i}\in\mathbb{Z}\right\} \label{3.3c}\end{equation}
which is also called the coweight lattice of $G$ and which has the
root lattice of the dual group $G^{\vee}$(or coroot lattice of $G$)\begin{equation}
\Lambda_{r}(G^{\vee})=\left\{ \beta=\sum_{i=1}^{r}n_{i}\alpha_{i}^{\vee},\,\,\,\,\,\,\, n_{i}\in\mathbb{Z}\right\} \label{3.3d}\end{equation}
as subset.

Let \[
I_{ij}=2\delta_{ij}-K_{ij}\]
 which is called the incident matrix. One can show that the eigenvalues
of $I_{ij}$ are \[
\lambda(\nu)=2\cos\frac{\pi\nu}{h}\]
where $h$ is the Coxeter number of $\mathfrak{g}$ and $\nu$ are
the exponents of $\mathfrak{g}$. For $\mathfrak{g}=su(n)$, the exponents
are $\nu=1,2,\,...,\, n-1$ and $h=n$. 

Let $x_{i}^{(\nu)}$ be the left eigenvector of $I_{ij}$ associated
to the eigenvalue $\lambda(\nu)$. Then \[
y_{i}^{(\nu)}=\alpha_{i}^{2}x_{i}^{(\nu)}/2\]
 is the right eigenvector of $I_{ij}$ with same eigenvalue $\lambda(\nu)$.
We shall adopt the normalization\[
x_{i}^{(\mu)}y_{i}^{(\nu)}=\delta_{\mu\nu}.\]
 Clearly $x^{(\mu)}$and $y^{(\mu)}$are also eigenvectors of the
Cartan matrix with\begin{equation}
K_{ij}y_{j}^{(\nu)}=2\left(1-\cos\frac{\pi\nu}{h}\right)y_{i}^{(\nu)}=4\sin^{2}\frac{\pi\nu}{2h}\, y_{i}^{(\nu)}.\label{3.3e}\end{equation}
The incident matrix $I_{ij}$ has strictly positive entries and it
is irreducible since we are considering $\mathfrak{g}$ simple. Therefore,
we can apply the Perron-Frobenius theorem which says that for a given
non-negative irreducible matrix $M$, there exist a eigenvalue $\mu$
such that $\mu\geq|\lambda|$ for all eigenvalues $\lambda$ of $M$
and this eigenvalue $\mu$ can be associated with strictly positive
left and right eigenvectors. Since for any algebra $\mathfrak{g}$,
$\nu=1$ is always the smallest exponent and $\nu=h-1$ is the largest
one, then \[
\lambda(1)=2\cos\frac{\pi}{h}\]
is the largest eigenvalue of $I_{ij}$ and we conclude that the corresponding
eigenvector components $x_{i}^{(1)}$ and $y_{i}^{(1)}$ never vanish
and can be taken positive. The other eigenvectors necessarily have
negative components. Therefore we call $x_{i}^{(1)}$ ($y_{i}^{(1)}$)
the components of the left (right) Perron-Frobenius eigenvector of
$I_{ij}$ and $K_{ij}$. Some of these (non-normalised) vectors are
listed in table 2, using the Dynkin diagram numbering convention of
table 1.

\begin{table}
\begin{center}\begin{tabular}{|c|c|c|}
\hline 
$SU(n+1)$ $(h=n+1)$&
$Spin(2n)\,(h=2n-2)$&
$E_{6}\,(h=12)$\tabularnewline
\hline
\hline 
$\begin{array}{c}
x_{1}^{(1)}=\sin(\pi/h)\\
x_{2}^{(1)}=\sin(2\pi/h)\\
x_{3}^{(1)}=\sin(3\pi/h)\\
\vdots\\
x_{n-1}^{(1)}=\sin[(n-1)\pi/h]\\
x_{n}^{(1)}=\sin(n\pi/h)\end{array}$&
$\begin{array}{c}
x_{1}^{(1)}=\sin(\pi/h)\\
x_{2}^{(1)}=\sin(2\pi/h)\\
\vdots\\
x_{n-2}^{(1)}=\sin[(n-2)\pi/h]\\
x_{n-1}^{(1)}=1/2\\
x_{n}^{(1)}=1/2\end{array}$ &
$\begin{array}{c}
x_{1}^{(1)}=\sin(\pi/h)\\
x_{2}^{(1)}=\sin(2\pi/h)\\
x_{3}^{(1)}=\sin(3\pi/h)\\
x_{4}^{(1)}=\sin(2\pi/h)\\
x_{5}^{(1)}=\sin(\pi/h)\\
x_{6}^{(1)}=\sin(8\pi/h)-\sin(2\pi/h)\end{array}$\tabularnewline
\hline
\end{tabular}\end{center}

\caption{\label{Perron}Some (non-normalised) left Perron-Frobenius eigenvectors
$x_{i}^{(1)}$ and Coxeter numbers $h$.}
\end{table}

\section{Vacuum solutions }

Let us now analyze some vacuum solutions of our theory. A vacuum solution
$\phi^{\textrm{vac}}$ breaks the gauge group $G$ to a subgroup $G_{\phi}$
which consist of the group elements which commutes with $\phi^{\textrm{vac}}$.
Considering that $\phi^{\textrm{vac}}$ can be embedded in a Cartan
subalgebra,  we can write \begin{equation}
\phi^{\textrm{vac}}=v\cdot H\label{4.1}\end{equation}
where $v$ is a $r$ component real vector which can be expanded in
the basis of the fundamental coweight vectors\begin{equation}
v=v_{i}\lambda_{i}^{\vee}.\label{4.2}\end{equation}
 If all coefficients $v_{i}$ do not vanish, then $G$ is broken to
the maximal torus $U(1)^{r}$ \cite{go review} associated to the
CSA $\mathfrak{h}$ which corresponds to the first symmetry breaking
in (\ref{1}) . The compact $U(1)$ factors are associated to the
group elements%
\footnote{No summation is assumed for the index $i$.%
}\begin{equation}
\exp\left\{ 2\pi ia^{(i)}\lambda_{i}^{\vee}\cdot H\right\} ,\,\,\,\,\,\,\,\, i=1,\,2,\,....,\, r\label{4.2a}\end{equation}
where $a^{(i)}$ are real parameters . 

In \cite{k2003} we considered a vacuum solution of the form\begin{eqnarray}
\phi_{1}^{\textrm{vac}} & \propto & \delta\cdot H\,,\,\,\,\,\,\,\delta\equiv\sum_{j=1}^{r}\lambda_{j}^{\vee}=\frac{1}{2}\sum_{\alpha>0}\alpha^{\vee},\label{4.3}\\
\phi_{2}^{\textrm{vac}} & \propto & \sum_{i=1}^{r}\sqrt{c_{i}}E_{-\alpha_{i}}\,,\,\,\,\,\,\,\, c_{i}=\sum_{j=1}^{r}\left(K_{ij}^{-1}\right)=\lambda_{i}\cdot\delta.\nonumber \end{eqnarray}
where $\delta$ is the dual Weyl vector. There was also a third complex
field in order to the theory be supersymmetric which did not produce
any extra symmetry breaking. This vacuum configuration produces the
symmetry breaking (\ref{1}). This result follows from the fact that
since $\phi_{1}^{\textrm{vac}}$ belongs to the Cartan subalgebra
with all the coefficients of $\lambda_{i}^{\vee}$ not vanishing,
it produces the first symmetry breaking in (\ref{1}). Then, as \[
\exp\left\{ 2\pi ia^{(i)}\lambda_{i}^{\vee}\cdot H\right\} E_{-\alpha_{j}}\exp\left\{ -2\pi ia^{(i)}\lambda_{i}^{\vee}\cdot H\right\} =\exp\left\{ -2\pi ia^{(i)}\delta_{ij}\right\} E_{-\alpha_{j}},\]
 the $U(1)$ group elements (\ref{4.2a}) will only commute with $\phi_{2}^{\textrm{vac}}$
if the constants $a^{(i)}$ are integers. Remembering that the center
$C_{G}$ of a group $G$, is formed by group elements\begin{equation}
\exp{2\pi i\omega\cdot H}\label{3.4}\end{equation}
where $\omega$ is a vector of the coweight lattice $\Lambda_{w}(G^{\vee})$,
we can conclude that $\phi_{2}^{\textrm{vac}}$ only commutes with
the center elements (\ref{3.4}) and produces the second symmetry
breaking in (\ref{1}). We showed \cite{k2003} that for this vacuum
configuration, the tensions of the BPS strings satisfies the Casimir
scaling when $G=SU(N)$. Let us now analyze some vacuum solutions
which produce the same symmetry breaking (\ref{1}). 

From the above example we can conclude that in order to produce the
symmetry breaking (\ref{1}) we can consider a general vacuum solution\begin{eqnarray}
\phi_{1}^{\textrm{vac}} & = & v\cdot H\,,\,\,\,\,\,\,\,\,\,\,\,\, v=v_{i}\lambda_{i}^{\vee},\label{3.18}\\
\phi_{2}^{\textrm{vac}} & = & \sum_{l=1}^{r}b_{l}E_{-\alpha_{l}},\nonumber \end{eqnarray}
where $v_{i}$ are non vanishing real constants and $b_{l}$ must
be non vanishing complex constants in order to $G$ to be broken to
$C_{G}$.

The vacua of our theory are solutions of\[
G_{\mu\nu}=D_{\mu}\phi_{s}=V(\phi,\phi^{*})=0.\]
The condition $V(\phi,\phi^{*})=0$ for the potential (\ref{8}) implies
that \[
\left[\phi_{1}^{\dagger},\phi_{1}\right]+\left[\phi_{2}^{\dagger},\phi_{2}\right]=m\textrm{Re}\left(\phi_{1}\right).\]
 Using the configuration (\ref{3.18}) in this condition it results
\begin{equation}
m\left(K^{-1}\right)_{ij}v_{j}=\left|b_{i}\right|^{2}.\label{4.11}\end{equation}
 From this equation, we conclude that when $m=0$, then $b_{i}=0$
and $\phi_{2}^{\textrm{vac}}=0$, and it happens the first symmetry
breaking in (\ref{1}), which corresponds to the Coulomb phase. In
order to happen the second symmetry breaking, all components $v_{i}$
and $b_{i}$ must be non vanishing, and therefore we must have $m\neq0$,
which means that the term $X_{a}$, given by Eq. (\ref{5e}), must
not vanish. 

In principle, Eq. (\ref{4.11}) can has various solutions depending
on $G$. However there are two which hold for any $G$. One solution
is \begin{equation}
v_{i}=a\,,\,\,\,\,\,\, b_{i}=\sqrt{am\sum_{j=1}^{r}\left(K^{-1}\right)_{ij}},\label{4.12}\end{equation}
where $a$ is a positive real constant. This solution gives rise to
the vacuum (\ref{4.3}).

The other solution is to consider that $v_{j}$ are the components
of a right eigenvector of the Cartan matrix $K_{ij}$. However, the
components $v_{j}$ can not vanish and from the relation (\ref{4.11})
we also see that they can not be negative since $m$ and the eigenvalues
of $K_{ij}$ are positive. Therefore, from the discussion in the previous
section we can conclude that $v_{j}$ can only be proportional to
the Perron-Frobenius right eigenvector of $K_{ij}$. Hence,\begin{equation}
v_{i}=ay_{i}^{(1)}\,,\,\,\,\,\,\,\, b_{i}=\frac{1}{2\sin\frac{\pi}{2h}}\sqrt{amy_{i}^{(1)}}\label{4.13}\end{equation}
 is a solution where $a$ is a positive constant and the corresponding
vacuum solution \begin{eqnarray}
\phi_{1}^{\textrm{vac}} & = & a\sum_{i=1}^{r}y_{i}^{(1)}\lambda_{i}^{\vee}\cdot H,\label{3.5a}\\
\phi_{2}^{\textrm{vac}} & = & \frac{\sqrt{am}}{2\sin\frac{\pi}{2h}}\sum_{i=1}^{r}\sqrt{y_{i}^{(1)}}E_{-\alpha_{i}},\nonumber \end{eqnarray}
also produces the symmetry breaking (\ref{1}). This vacuum is very
interesting since it gives rise to the sine law scaling for the ratios
of BPS string tensions and a possible connection with Affine Toda
Field theories as we shall see in the next sections.

\section{The Coulomb phase}

Let us analyze the Coulomb phase which happens when $m=0$. In this
phase, there exist free monopoles. For a symmetry breaking produced
by an arbitrary vacuum configuration (\ref{3.18}), with $v_{i}\neq0$
and $b_{i}=0$, one can construct monopole solutions for each root
$\alpha$ such that $\alpha\cdot v\neq0$, which has magnetic charge
\cite{bais} 

\begin{equation}
g_{\alpha}=\frac{1}{\left|\phi_{1}^{\textrm{vac}}\right|}\oint d^{2}S_{i}\left[Re\left(\phi_{1}\right)_{a}B_{a}^{i}\right]=\frac{2\pi}{e}\frac{v\cdot\alpha^{\vee}}{|v|}.\label{4.10}\end{equation}
Since in our case the scalar product $\alpha\cdot v$ with any root
$\alpha$ never vanishes, we can then construct monopoles for any
root $\alpha$. The vacuum solution $\phi_{1}^{\textrm{vac}}=v\cdot H$
singles out a particular $U(1)$ direction which we call $U(1)_{v}$.
This magnetic charge is equal to the monopole magnetic flux in this
$U(1)_{v}$ direction.

The mass of the BPS monopole associated to a root $\alpha$ is \cite{bais}\[
M_{\alpha}^{\textrm{mon}}=\frac{4\pi}{e}|v\cdot\alpha^{\vee}|.\]
However, the stable or fundamental BPS monopoles are the ones associated
to the simple roots $\alpha_{i}$ \cite{eweinberg}. In particular,
for the vacuum (\ref{4.13}), the vector $v$ (\ref{3.18}) can be
written as\begin{equation}
v=ay_{i}^{(1)}\lambda_{i}^{\vee}=ax_{i}^{(1)}\lambda_{i}=\frac{a}{4\sin^{2}(\pi/2h)}x_{i}^{(1)}\alpha_{i},\label{4.4}\end{equation}
where in the last equality was used Eqs. (\ref{3.3}) and (\ref{3.3e}).
Therefore, for this vacuum, the masses for the stable BPS monopoles
are \begin{equation}
M_{\alpha_{i}}^{\textrm{mon}}=\frac{4\pi}{e}|v\cdot\alpha_{i}^{\vee}|=\frac{4\pi a}{e}x_{i}^{(1)}\,,\,\,\,\,\, i=1,\,2,\,...,\, r.\label{4.4a}\end{equation}
 Likewise, for each root $\alpha$ there is a massive gauge particle
associated to the step operator $E_{\alpha}$, but the stable massive
particles are the ones associated to the simple roots $\alpha_{i}$
with masses \cite{bais}\begin{equation}
M_{\alpha_{i}}^{\textrm{W}}=e|v\cdot\alpha_{i}|=aey_{i}^{(1)}\,,\,\,\,\,\, i=1,\,2,\,...,\, r.\label{4.4b}\end{equation}

Let us now see that this spectrum of masses of stable massive particles
and monopoles coincide with the spectrum of masses of particles and
solitons of affine Toda field theoryies (ATFT).

Affine Toda field theories are two dimensional integrable theories.
For each affine Lie algebra $\widehat{\mathfrak{g}}$ we can associated
an ATFT. For simplicity only the untwisted case will be considered.
They have $r$ scalar fields $\phi_{i}$ where $r$ is the rank of
the Lie algebra $\mathfrak{g}$, from which $\widehat{\mathfrak{g}}$
is constructed. Their Lagrangian can be written as\[
L=\frac{1}{2}\partial_{\mu}\phi\cdot\partial^{\mu}\phi-\mu^{2}\sum_{i=0}^{r}n_{i}e^{\beta\alpha_{i}\cdot\phi}\]
 where $\mu$ is a mass parameter and $\beta$ is an adimensional
constant. The scalar product is defined in the $r$ dimensional space
of fields $\phi_{i}$ and the integers $n_{i}$ are defined from the
expansion of the highest root $\psi$ in the basis of simple roots:
\begin{equation}
\psi=\sum_{i=1}^{r}n_{i}\alpha_{i}.\label{4.7}\end{equation}
 Moreover, we are considering that $\alpha_{0}=-\psi$ and $n_{0}=1$.
In particular, for $\mathfrak{\mathfrak{g}}=su(n)$, $n_{i}=1$, for
$i=1,\,2,\,...,\, n-1$. 

We shall consider that $\beta$ is imaginary which implies that the
theory has degenerated vacuum and solitons interpolating these vacua.
In this case, the theory has $r$ species of particles and $r$ species
of solitons \cite{HollowoodATFT}\cite{DioTurokUnder1}, one for each
node of the Dynkin diagram of $\mathfrak{g}$. The particle masses
are \cite{CorriganDorey}\cite{freeman}\cite{FringOliveLiao} \begin{equation}
M_{i}^{\textrm{part}}=\mu|\beta|\sqrt{2h}y_{i}^{(1)}\,,\,\,\,\, i=1,\,2,\,...,\, r\label{4.7a}\end{equation}
here $h$ is the Coxeter number of $\mathfrak{g}$. The soliton masses
are \cite{HollowoodATFT}\cite{DioTurokUnder1}\begin{equation}
M_{i}^{\textrm{sol}}=\frac{2h}{|\beta|^{2}}\frac{2}{\alpha_{i}^{2}}M_{i}^{\textrm{part}}=\frac{\mu(2h)^{3/2}}{|\beta|}x_{i}^{(1)}\,,\,\,\,\, i=1,\,2,\,...,\, r.\label{4.7b}\end{equation}
Each soliton species may have many topological charges, with same
masses.

One can easily check that the spectrum of masses of particles and
solitons of an ATFT associated to the affine algebra $\widehat{\mathfrak{g}}$
coincide respectively with the spectrum of masses of stable massive
gauge particles and BPS monopoles of our theory in the Coulomb phase
with gauge group associated to the algebra $\mathfrak{g}$, if the
couplings of the two theories satisfy the relations\begin{eqnarray}
\frac{e^{2}}{4\pi} & = & \frac{|\beta|^{2}}{2h},\label{4.20a}\\
a & = & \frac{\mu h}{\sqrt{\pi}}.\label{4.20b}\end{eqnarray}
 Note that this result holds also for Yang-Mills theories with gauge
groups $E_{8}$, $F_{4}$ and $G_{2}$ which, although do not have
stable $Z_{N}$ strings in the Higgs phase, they have monopoles in
the Coloumb phase. This result indicates a possible relation or {}``duality''
between ATFTs in two dimensions and Yang-Mills-Higgs in four dimensions
with the vacuum \begin{equation}
\phi_{1}^{\textrm{vac}}=a\sum_{i=1}^{r}y_{i}^{(1)}\lambda_{i}^{\vee}\cdot H\,\,,\,\,\,\,\,\phi_{2}^{\textrm{vac}}=0.\label{4.20c}\end{equation}
 One must observe that all these mass spectra we mentioned are at
the classical level. Therefore these possible relations probably only
holds exactly (i.e., at the quantum level) when these theories are
embedded in supersymmetric theories, as usual. Since $\phi_{1}^{\textrm{vac}}$
is in Cartan subalgebra, it is direct to see that the field configuration
(\ref{4.20c}), together with an extra field $\phi_{3}^{\textrm{vac}}=0$,
is a vacuum solution of the bosonic part of the ${\cal {\cal N}}=4$
potential\[
V=\frac{1}{2}\textrm{Tr}\left[\frac{e}{2}\sum_{s=1}^{3}\left(\left[\phi_{s}^{*},\phi_{s}\right]\right)\right]^{2}\]
and therefore gives rise to the same mass spectrum (\ref{4.4a}) and
(\ref{4.4b}) for the gauge particles and BPS monopoles in ${\cal {\cal N}}=4$
superYang-Mills theoreis. 

It is interesting to note that in \cite{DoreyHollowoodTong} it was
also observed a relation between BPS mass spectra for some two and
four dimensional theories. On the other hand, a relation between non-Abelian
monopoles and conformal invariant Toda theory was shown in \cite{Saveliev}\cite{GanoulisGodOlive}.
In those works it was shown that for a particular spherically symmetric
BPS monopole associated to the maximal $SU(2)$ subalgebra, $T_{3}=\delta\cdot H$,
$T_{\pm}=\sum_{i=1}^{r}\sqrt{\delta\cdot\lambda_{i}}E_{\pm\alpha_{i}}$
(like the vacuum configuration (\ref{4.3})), the monopole\'{ }s
radial function satisfy the equation of motion of conformal Toda field
theory .

Note that our theory in the Coulomb phase when embedded in a ${\cal {\cal N}}=4$
super Yang-Mills theory should satisfy the Montonen-Olive duality\cite{MontoneOlive},
with the monopoles and particles of the theory with gauge group $G$
and coupling $e$ being mapped respectively to the particles and monopoles
of the theory with gauge group $G^{\vee}$ and coupling $4\pi/e$.
Therefore, combining the above duality with Montonen-Olive duality
should imply a duality between ATFT associated to $\widehat{\mathfrak{g}}$
with coupling constants $(\mu,\beta)$ and ATFT associated to $\widehat{\mathfrak{g^{\vee}}}$
with coupling constants $(\mu,2h/\beta)$. This is consistent the
classical spectrum of masses of these theories. One can see that fact
remembering that if $K_{ij}$ is the Cartan matrix associated to the
algebra $\mathfrak{g}$ then the transposed $(K^{\textrm{T}})_{ij}$
is the Cartan matrix of the dual algebra $\mathfrak{g}^{\vee}$ and
the right (left) vectors of $K_{ij}$ are left (right) vectors of
$(K^{T})_{ij}$. Therefore the mass spectrum for particles (solitons)
of the ATFT associated to $\widehat{\mathfrak{g}}$ with coupling
constants $(\mu,\beta)$ is the same as the mass spectrum for solitons
(particles) of the ATFT associated to $\widehat{\mathfrak{g^{\vee}}}$
with coupling constants $(\mu,2h/\beta)$. However one must have in
mind that each soliton species have many different topological charges.
Therefore, similarly to the Yang-Mills theory in four dimensions,
this duality probably should hold only when ATFT is embed in a supersymmetric
theory where the number of particles is increased.

\section{The Higgs phase: the sine law and the Casimir scaling}

When $m\neq0$, $G$ is broken to its center $C_{G}$, which corresponds
to the Higgs phase and there exist $Z_{N}$ string solutions. In \cite{k2003}
we analyzed many properties of these solutions for the vacuum given
by Eq. (\ref{4.3}). Let us extend these results for a general vacuum
configuration given by Eq. (\ref{3.18}) which breaks $G$ to its
center $C_{G}$. In order to have finite string tension, asymptotically
this solutions have the form\begin{eqnarray}
\phi_{s}(\varphi,\rho\rightarrow\infty) & = & g(\varphi)\phi_{s}^{\textrm{vac}}g(\varphi)^{-1},\,\,\, s=1,\,2,\label{5.1}\\
W_{I}(\varphi,\rho\rightarrow\infty) & = & -\frac{1}{ie}\left(\partial_{I}g(\varphi)\right)g(\varphi)^{-1},\nonumber \end{eqnarray}
where $\phi_{s}^{\textrm{vac}}$ are the vacuum solutions (\ref{3.18}),
$\rho$ and $\varphi$ are the radial and angular coordinates and
the capital Latin letters $I,\, J$ denote the coordinates 1 and 2
orthogonal to the string. In order for the configuration to be single
valued, $g(\varphi+2\pi)g(\varphi)^{-1}\in C_{G}$. Considering\[
g(\varphi)=\exp i\varphi M,\]
 where $M$ is a generator of $\mathfrak{g}$, it results that $\exp2\pi iM\in C_{G}$.
From this condition we can consider\[
M=\omega\cdot H\]
 with $\omega\in\Lambda_{w}(G^{\vee})$. Then, the asymptotic form
of the $Z_{N}$ string solution can be written as \begin{eqnarray}
\phi_{1}(\varphi,\rho\rightarrow\infty) & = & v\cdot H,\nonumber \\
\phi_{2}(\varphi,\rho\rightarrow\infty) & = & \sum_{i=1}^{r}b_{i}\left\{ \exp\left(-i\varphi\omega\cdot\alpha_{i}\right)\right\} E_{-\alpha_{i}},\label{5.4}\\
W_{I}(\varphi,\rho\rightarrow\infty) & = & \frac{\epsilon_{IJ}x^{J}}{e\rho^{2}}\omega\cdot H\,\,,\,\,\, I=1,2.\nonumber \end{eqnarray}
One can see that for each element $\omega$ in the coweight lattice
$\Lambda_{w}(G^{\vee})$ we can associate a string solution. Let us
review how  the $Z_{N}$ string solutions are associated to distinct
topological sectors of $\Pi_{1}(G/C_{G})$\cite{k2003} for a general
group $G$. In order to do that we must remember that since $\Lambda_{r}(G^{\vee})$
is a sublattice (or subgroup) of $\Lambda_{w}(G^{\vee})$, we can
define the quotient $\Lambda_{w}(G^{\vee})/\Lambda_{r}(G^{\vee})$
by identifying points $\Lambda_{w}(G^{\vee})$ which differ by an
element of the coroot lattice $\Lambda_{r}(G^{\vee}).$ In \cite{gno},
it was showed that \begin{equation}
C_{G}\simeq\Lambda_{w}(G^{\vee})/\Lambda_{r}(G^{\vee}).\label{5.9}\end{equation}
 On the other hand, as explained in detail in \cite{OliveTurokdynkinsymmetry},
the center group $C_{G}$ is isomorphic to the symmetry group $W_{0}$
of the extended Dynkin diagram formed by the transformations $\tau$
where the node $0$ is not fixed, but mapped to another node $j=\tau(0)$.
The elements of $W_{0}$ may be labelled by those nodes symmetrically
related to the node $0$, as shown in table 1, and are represented
by black nodes in the extended Dynkin diagrams. As a consequence,
the quotient (\ref{5.9}) can be represented by the cosets \begin{equation}
\Lambda_{r}(G^{\vee}),\,\,\,\,\,\lambda_{\tau(0)}^{\vee}+\Lambda_{r}(G^{\vee}),\,\,\,\,\,\lambda_{\tau^{2}(0)}^{\vee}+\Lambda_{r}(G^{\vee}),\,...,\,\,\,\,\lambda_{\tau^{n}(0)}^{\vee}+\Lambda_{r}(G^{\vee})\label{3.6}\end{equation}
where the weights $\lambda_{\tau^{q}(0)}$are associated to nodes
related to the node $0$ by a symmetry transformation. Such weights
are called the minimal weights of $\mathfrak{g}$. A fundamental weight
$\lambda_{k}$ is minimal if $\lambda_{k}^{\vee}\cdot\psi=1$, where
$\psi$ is the highest root (\ref{4.7}). One can lift $\tau$ to
an automorphism of the Lie algebra $\mathfrak{g}$ and show that the
center group elements (\ref{3.4}), with $\omega$ belonging to a
given coset in (\ref{3.6}) are associated to the same center element
of $C_{G}$\cite{OliveTurokUnder2}. In other words, the coweights
in a coset are associated to the same center element. The representations
of $G^{\vee}$ with these weights are said to be in the same N-ality.
When $\omega$ belongs to $\Lambda_{r}(G^{\vee})$, the group element
(\ref{3.4}) is the identity since when it acts on any weight state
$\left|\lambda\right\rangle $ of any representation of $G$,\[
\exp\left\{ 2\pi i\omega\cdot H\right\} \left|\lambda\right\rangle =\exp\left\{ 2\pi i\omega\cdot\lambda\right\} \left|\lambda\right\rangle =\left|\lambda\right\rangle ,\]
using the fact that $\omega\in\Lambda_{r}(G^{\vee}),$ $\lambda\in\Lambda_{w}(G)$
and the orthonormality condition (\ref{3.3a}).

Since the topological sectors of the strings solutions are given by
\[
\Pi_{1}(G/C_{G})=C_{G},\]
we can conclude that the $Z_{N}$ string solutions (\ref{5.4}) are
separated in topological sectors according to the coset (\ref{3.6})
which $\omega$ belongs\cite{k2003}. When $\omega$ belongs to $\Lambda_{r}(G^{\vee})$,
the group element (\ref{3.4}) is the identity, and the corresponding
string solution is in the trivial topological sector. 

Let us analyze for example how the string solutions are split in topological
sectors for the groups $E_{6},\, Spin(2n)$ and $SU(n)$ which are
the groups in which the center groups have order greater than two,
which are the most interesting. All these groups are simply laced,
i.e. $\alpha_{i}^{\vee}=\alpha_{i}$, $\lambda_{i}^{\vee}=\lambda_{i}$,
$G^{\vee}=G$ and so the weight and the coweight lattice are the same.
In table 1 are listed the elements of $W_{0}$, from which we obtain
how the weight lattice split in cosets.

For $G=E_{6}$, the weight lattice split in the three cosets\begin{equation}
\Lambda_{r}(E_{6}),\,\,\,\lambda_{1}+\Lambda_{r}(E_{6}),\,\,\,\lambda_{5}+\Lambda_{r}(E_{6}),\label{5.11}\end{equation}
 and the group elements (\ref{3.4}) with $\omega$ belonging to each
of these three cosets are associated to the three elements of $Z_{3}$,
the center of $E_{6}$. $\lambda_{1}$ and $\lambda_{5}$ are the
highest weights of the representations $27$ and $\overline{27}$.

For $G=Spin(2n),$ the universal covering group of $SO(2n)$, with
$n\geq4$, the weight lattice split in the four cosets,\begin{equation}
\Lambda_{r}(Spin(2n)),\,\,\,\lambda_{1}+\Lambda_{r}(Spin(2n)),\,\,\,\lambda_{n-1}+\Lambda_{r}(Spin(2n)),\,\,\,\lambda_{n}+\Lambda_{r}(Spin(2n)),\label{5.12}\end{equation}
 where $\lambda_{1}$ is the highest weight of the $n$ dimensional
vector representation and $\lambda_{2n-1}$ and $\lambda_{2n}$ are
the highest weights of the spinor representations of $Spin(2n)$.
Then, the group elements (\ref{3.4}) with $\omega$ belonging to
each of these cosets are associated to the four elements of the center
group of $Spin(2n)$, $Z_{2}\times Z_{2}$ when $n$ is even or $Z_{4}$
when $n$ is odd. 

For $G=SU(n)$, the weight lattice split in $n$ cosets,\begin{equation}
\Lambda_{r}(SU(n)),\,\,\,\lambda_{1}+\Lambda_{r}(SU(n)),\,\,\,\lambda_{2}+\Lambda_{r}(SU(n)),\,\,\,.....,\,\,\,\lambda_{n-1}+\Lambda_{r}(SU(n)),\label{5.13}\end{equation}
 where $\lambda_{k}$ is the fundamental weight associated to the
representation which is the antisymmetric tensor product of $k$ $n$-dimensional
fundamental representations. The group elements (\ref{3.4}) with
$\omega$ belonging to each of these $n$ cosets are associated to
the $n$ elements of $Z_{n}$. One can see this result explicitly
by acting these group elements on the $n$ weight states\[
\left|\lambda_{1}\right\rangle \,,\,\,\,\,\left|\lambda_{1}-\sum_{i=1}^{k}\alpha_{i}\right\rangle ,\,\,\,\,\, k=1,2,\,\,....,n-1\]
of the $n$ dimensional representation of $SU(n)$, which results\begin{eqnarray*}
\exp\left\{ 2\pi i\left[\lambda_{m}+\Lambda_{r}(SU(n))\right]\cdot H\right\} \left|\lambda_{1}-\sum_{i=1}^{k}\alpha_{i}\right\rangle  & = & \exp\left\{ 2\pi i\lambda_{m}\cdot\lambda_{1}\right\} \left|\lambda_{1}-\sum_{i=1}^{k}\alpha_{i}\right\rangle \\
 & = & \exp\left\{ 2\pi i\frac{m}{n}\right\} \left|\lambda_{1}-\sum_{i=1}^{k}\alpha_{i}\right\rangle .\end{eqnarray*}
The same result holds when acting on $\left|\lambda_{1}\right\rangle $.

From the string asymptotic form (\ref{5.4}) one can propose an ansatz
for all the fields in the whole space. However, for us it is only
important to consider the ansatz that $\phi_{1}(\varphi,\rho)$ is
constant in the whole space and equal to its asymptotic value, i.e.,\begin{equation}
\phi_{1}(\varphi,\rho)=v\cdot H.\label{5.17}\end{equation}
 In particular for the BPS strings, one can deduce this configuration
from the BPS equation $D_{\mp}\phi_{1}=0$ and asymptotic form (\ref{5.4})
\cite{k2003}. The construction of the $Z_{N}$ string solution for
the whole space is analyzed in \cite{integrability Z(N) strings}.

For the string associated to a vector $\omega$ given by Eqs. (\ref{5.4})
and (\ref{5.17}), the string flux (\ref{7b}) is \begin{equation}
\Phi_{\textrm{st}}=\frac{1}{\left|\phi_{1}^{\textrm{vac}}\right|}\int d^{2}x\left[Re\left(\phi_{1}\right)_{a}B_{3a}\right]=-\frac{1}{|v|}\oint dl_{I}\textrm{Tr}\left[v\cdot H\, W_{I}\right]=\frac{2\pi}{e}\frac{v\cdot\omega}{|v|}.\label{5.7}\end{equation}
Similarly to the monopole flux (\ref{4.10}), this is the string flux
in the $U(1)_{v}$ direction generated by $v\cdot H$. 

In the Higgs phase, the monopoles magnetic lines can not spread radially
over space. However, since any coroot $\alpha^{\vee}$ can be expanded
as an integer linear combination of coweights, we can conclude that
any monopole flux (\ref{4.10}) can be always as an integer linear
combination of string fluxes (\ref{5.7}). Note that this result holds
for an arbitrary vacuum (\ref{3.18}) with non-vanishing $v_{i}$,
extending therefore our previous result. Then, the monopole magnetic
lines form a set of $Z_{N}$ strings and monopoles become confined
as analyzed in detail in \cite{k2003}. Note that this flux matching
happens not only with respect to $v\cdot H$, which is the generator
of $U(1)_{v}$, but for any other Cartan generator $H_{i}$, $i=1,\,2,\,...,\, r$.
Then a pair of monopole-antimonopole associated to a coroot $\alpha^{\vee}$
would be confined by a string associated to $\omega=\alpha^{\vee}$.

From the string flux (\ref{5.7}) we can obtain that the lower bound
for the tension (\ref{7a}) for a string associated to $\omega=\lambda_{k}^{\vee}-\beta^{\vee}$,
with $\beta^{\vee}\in\Lambda_{r}(G^{\vee})$, is

\begin{equation}
T_{\omega}\geq\pi m\left|v\cdot\omega\right|=\pi m\left|v\cdot\left(\lambda_{k}^{\vee}-\beta^{\vee}\right)\right|.\label{5.8}\end{equation}
The bound holds for the BPS strings. We shall only consider here the
BPS strings.

In $QCD$ with gauge group $SU(N)$, it is believed that the chromoelectric
flux tubes carries charge in the center $Z_{N}$ of the gauge group,
but their tension in general could depend on the representation of
the sources \cite{greensite}. However, it is believed that for long
enough strings it become energetically favorable a pair of gluons
to pop out to bind with the quark and antiquark charges. For all representations
associated to the same center element (i.e. in the same $N$-ality
of $SU(N)$), the energetically most favorable representation of the
quark-gluon bound state will be the lowest dimensional representation.
There are mainly two conjectures for the ratios of these asymptotic
tensions: the Casimir scaling and the sine law scaling. 

As we have seen, the $Z_{N}$ strings associated to the same center
element are those with $\omega$ in the same coset. In general, they
do not have same tensions as can be seen in Eq. (\ref{5.8}). For
a long enough string associated to $\omega$, it could pop out a pair
of monopole-antimonopole confined by a string associated to a coroot
$\alpha^{\vee}$as described above and the string associated to $\omega$
would decay to a string associated to $\omega-\alpha^{\vee}$ which
clearly is in the same coset as $\omega$ and therefore associated
to the same center element. From the monopole mass (\ref{4.4a}) and
string tension bound (\ref{5.8}) we obtain that the threshold length
$l^{\textrm{th}}$for this decay to happen is\[
l^{\textrm{th}}=\frac{2M_{\alpha}^{\textrm{mon}}}{T_{\alpha^{\vee}}}\leq\frac{8}{em}.\]
The $Z_{N}$ strings can have different tensions for weights in a
same representation. Therefore, in order to compare with results of
QCD strings we shall associated to a representation the tension of
its hightest weight. In order to determine the smallest tension in
the same topological sector it is convenient to write the vector $v$
(\ref{4.2}) in the simple root basis:\[
v=u_{i}\alpha_{i},\,\,\,\,\, u_{i}=\frac{2}{\alpha_{i}^{2}}\left(K^{-1}\right)_{ij}v_{j}\]
 where all the entries of $K^{-1}$ are positive. A highest coweight
can be written as $\omega=p_{i}\lambda_{i}^{\vee}$ where $p_{i}$
are integers and $p_{i}\geq0$, and the tension (\ref{5.8}) of the
BPS string associated to $\omega$ can be written as \[
T_{\omega}=\pi mu_{i}p_{i}.\]
 For the vacua and gauge groups we are considering below, one can
check that the highest weight associated to the smallest tension for
each topological sector, is the the minimal coweight $\lambda_{\tau^{q}(0)}^{\vee}$.
We shall call minimal strings the strings associated to minimal coweights.
From Eq. (\ref{5.8}) we see that their tensions are 

\begin{equation}
T_{\omega}=\pi m\left|v\cdot\lambda_{\tau^{q}(0)}^{\vee}\right|.\label{5.18}\end{equation}
For a theory with vacuum given by Eq. (\ref{4.3}), the ratio of BPS
minimal string tensions satisfy the Casimir scaling, for the gauge
group $G=SU(n)$\cite{k2003}. Let us verify that for the vacuum given
by Eq. (\ref{3.5a}), the ratios of tensions give rise to the sine
law scaling, when $G=SU(n)$. Let us also analyze how these scalings
generalize for other gauge groups. Note that since the groups considered
below are simply laced, therefore $\lambda_{k}^{\vee}=\lambda_{k}$.

\subsection{Sine law scaling}

For the vacuum with $v$ given by (\ref{4.4}), the BPS string tension
(\ref{5.18}) associated to $\omega=\lambda_{k}^{\vee}$ is \begin{equation}
T_{\lambda_{k}^{\vee}}=\frac{\pi ma}{4\sin^{2}\left(\pi/2h\right)}x_{k}^{(1)}.\label{5.19}\end{equation}
 Therefore, for this vacuum the tensions are proportional to the components
of the left Perron-Frobenius eigenvector $x_{k}^{(1)}$. From table
2 we obtain the following BPS minimal string tensions%
\footnote{In these results we absorbed a possible normalization constant of
$x_{i}^{(1)}$redefining the constant $a$. %
}:

\subsubsection*{a) For $G=SU(n)$}

For $SU(n)$, for each fundamental minimal weight $\lambda_{k}$,
$k=1,2,\,...,,n-1$, we associate a coset and hence a non-trivial
string topological sector. The corresponding BPS string tensions are\[
T_{\lambda_{k}}=\frac{\pi ma}{4\sin^{2}(\pi/2n)}\sin\frac{k\pi}{n},\,\,\,\, k=1,\,2,\,...,\, n-1.\]
 Taking the ratios of all tension with the smallest string tension
it results\[
\frac{T_{\lambda_{k}}}{T_{\lambda_{1}}}=\frac{\sin(k\pi/n)}{\sin(\pi/n)},\,\,\,\,\, k=1,2,\,....,\, n-1\]
which is exactly the sine law scaling.

\subsubsection*{b) For $G=Spin(2n)$, $n\ge4$}

For $Spin(2n)$, for each fundamental minimal weight $\lambda_{1},\,\lambda_{2n-1},\,\lambda_{2n}$,
we associate a coset and hence a non-trivial string topological sector.
The corresponding BPS string tensions associated to these weights
are\begin{eqnarray*}
T_{\lambda_{1}} & = & \frac{a\pi m\sin[\pi/2(n-1)]}{2\sin^{2}[\pi/4(n-1)]},\\
T_{\lambda_{2n-1}} & = & T_{\lambda_{2n}}=\frac{a\pi m}{4\sin^{2}[\pi/4(n-1)]}.\end{eqnarray*}
Note that for $n=4$, $T_{\lambda_{1}}=T_{\lambda_{2n-1}}=T_{\lambda_{2n}}$,
which is due to the symmetry of the $so(8)$ Dynkin diagram. Taking
the ratios of all tensions with the smallest string tension it results
\begin{equation}
\frac{T_{\lambda_{2n-1}}}{T_{\lambda_{1}}}=\frac{T_{\lambda_{2n}}}{T_{\lambda_{1}}}=\frac{1}{2\sin[\pi/2(n-1)]}.\label{5.25}\end{equation}
For large $n$, this ratio gives\begin{equation}
\frac{T_{\lambda_{2n-1}}}{T_{\lambda_{1}}}=\frac{T_{\lambda_{2n}}}{T_{\lambda_{1}}}\rightarrow\frac{n}{\pi}.\label{5.25a}\end{equation}

\subsection*{c) For $G=E_{6}$}

For each fundamental minimal weight $\lambda_{1}$ and $\lambda_{5}$
of $G=E_{6}$ are associated cosets. The BPS string tensions for these
weights are\[
T_{\lambda_{1}}=T_{\lambda_{5}}=\frac{\pi ma}{4\sin^{2}(\pi/24)}\sin\frac{\pi}{12}.\]

\bigskip{}
One can note that for a general gauge group, the string tension (\ref{5.19})
associated to $\lambda_{k}^{\vee}$ is proportional to $x_{k}^{(1)}$
and the topological sector is associated to the coset $\lambda_{k}^{\vee}+\Lambda_{r}(G^{\vee})$.
Similarly, in ATFTs, a soliton species associated to the $k^{\textrm{th}}$
node of the Dynkin diagram has mass proportional to $x_{k}^{(1)}$,
given by Eq. (\ref{4.7b}), and the topological charge has the form
$2\pi i(\lambda_{k}^{\vee}+\Lambda_{r}(G^{\vee}))/\beta$ \cite{HollowoodATFT}\cite{DioTurokUnder1}\cite{MackOlive1}.
Taking the ratios of the tensions of the BPS strings associated to
the fundamental coweights (\ref{5.19}) (not only the minimal ones)
for any gauge group $G$ we obtain\[
\frac{T_{\lambda_{i}^{\vee}}}{T_{\lambda_{k}^{\vee}}}=\frac{M_{i}^{\textrm{sol}}}{M_{k}^{\textrm{sol}}},\]
where $M_{i}^{\textrm{sol}}$ are the soliton masses (\ref{4.7b})
of the corresponding affine Toda field theory. Therefore, it may exist
some relation between these topological solutions. However, this possible
connection between $Z_{N}$ strings and solitons of ATFTs must yet
be clarified. In \cite{shifman} it was also shown that in $CP(N-1)$
sigma models, the tension between a $k$-kink and $k$-antikink also
satisfies the sine law scaling for the group $SU(N)$.

\subsection{Casimir scaling}

Let us now consider the vacuum (\ref{3.18}), (\ref{4.12}) with \[
v=a\sum_{i=1}^{r}\lambda_{i}^{\vee}=a\delta,\]
where $a$ is a real parameter. Then, from (\ref{5.18}) we have that
the tension of a BPS string for $\omega=\lambda_{k}^{\vee}$ is\begin{equation}
T_{\lambda_{k}^{\vee}}=a\pi m\lambda_{k}^{\vee}\cdot\delta.\label{5.29}\end{equation}
 This expression can be written\cite{k2003} in terms of the value
of the quadratic Casimir of a representation with fundamental weight
$\lambda_{k}^{\vee}$ of the dual Lie algebra $\mathfrak{g}^{\vee}$
\begin{equation}
C(\lambda_{k}^{\vee})=\lambda_{k}^{\vee}\cdot\left(\lambda_{k}^{\vee}+2\delta\right)\label{5.29a}\end{equation}
 as \[
T_{\lambda_{k}^{\vee}}=\frac{a\pi m}{2}\left(C(\lambda_{k}^{\vee})-\lambda_{k}^{\vee}\cdot\lambda_{k}^{\vee}\right).\]
Alternatively, from the definition of the dual Weyl vector $\delta$,
we can write \[
\lambda_{k}^{\vee}\cdot\delta=\frac{2}{\alpha_{k}^{2}}\sum_{i=1}^{r}\left(K^{-1}\right)_{ki}\]
where $\left(K^{-1}\right)_{ki}$ is tabulated in any standard Lie
algebra book. From this relation one can check explicitly, case by
case, that for any fundamental coweight $\lambda_{k}^{\vee}$ which
is minimal\begin{equation}
\lambda_{k}^{\vee}\cdot\delta=\frac{h}{2\left(h+1\right)}C(\lambda_{k}^{\vee})\label{5.29c}\end{equation}
 where $h$ is the Coxeter number of $G$ (which is also the Coxeter
number of $G^{\vee})$ given in Table 2. Therefore, the tension (\ref{5.29})
for a BPS string for minimal $\lambda_{k}^{\vee}$ can be writen as\begin{equation}
T_{\lambda_{k}^{\vee}}=a\pi m\frac{h}{2\left(h+1\right)}C(\lambda_{k}^{\vee})\label{5.29d}\end{equation}
 and the ratio of BPS string tensions associated to any minimal coweights
$\lambda_{k}^{\vee}$ and $\lambda_{j}^{\vee}$ can be written as
\begin{equation}
\frac{T_{\lambda_{k}^{\vee}}}{T_{\lambda_{j}^{\vee}}}=\frac{C(\lambda_{k}^{\vee})}{C(\lambda_{j}^{\vee})}\label{5.40}\end{equation}
which is a generalization of the Casimir scaling for any group $G$.
However, it is important to emphasize that (\ref{5.29d}) and (\ref{5.40})
hold only for minimal coweights $\lambda_{k}^{\vee}$, otherwise one
must use (\ref{5.29}).

The relation (\ref{5.29c}) can be proved in general in the following
way: any minimal weight $\lambda_{k}$ can be related to the node
$0$ of the extended Dynkin diagram by a symmetry transformation $\tau$.
For each of these transformations  \cite{OliveTurokUnder2},\[
\tau(\delta)-\delta=-h\lambda_{\tau(0)}^{\vee}.\]
 One can check this identity by taking scalar products with simple
roots. Moreover,\[
\tau(\delta)\cdot\lambda_{\tau(0)}^{\vee}=-\delta\cdot\lambda_{\tau(0)}^{\vee}.\]
Thus,\[
\delta\cdot\lambda_{\tau(0)}^{\vee}=\frac{h}{2}\lambda_{\tau(0)}^{\vee}\cdot\lambda_{\tau(0)}^{\vee}.\]
Therefore, for representations of with highest weight $\lambda_{k}^{\vee}$
minimal, the quadratic Casimir (\ref{5.29a}) can be written as\[
C(\lambda_{k}^{\vee})=2\left(\frac{h+1}{h}\right)\lambda_{k}^{\vee}\cdot\delta.\]

Let us now analyze the string tensions (\ref{5.29d}) and (\ref{5.40})
for some gauge groups.

\subsubsection*{a) For $G=SU(n)$}

For $G=SU(n)$, $h=n$ and \[
C(\lambda_{k})=\frac{\left(n+1\right)k\left(n-k\right)}{n}.\]
Therefore, for the minimal fundamental weights $\lambda_{k},\, k=1,\,2,\,...,\, n-1$,
the BPS string tensions are \[
T_{\lambda_{k}}=\frac{a\pi m}{2}k\left(n-k\right),\,\,\,\,\textrm{for }k=1,\,2,\,....,\, n-1,\]
 which results the Casimir scaling for the $Z_{N}$ BPS strings\begin{equation}
\frac{T_{\lambda_{k}}}{T_{\lambda_{1}}}=\frac{k\left(n-k\right)}{n-1},\label{5.30}\end{equation}
 obtained in \cite{k2003}.

\subsubsection*{b) For $G=Spin(2n)$, $n\geq4$ }

For $Spin(2n)$, $h=2n-2$ and , \begin{eqnarray*}
C(\lambda_{1}) & = & 2n-1,\\
C(\lambda_{n-1}) & = & C(\lambda_{n})=\frac{n\left(2n-1\right)}{4}.\end{eqnarray*}
Therefore, for the minimal fundamental weights $\lambda_{1}$, $\lambda_{n-1}$
and $\lambda_{n}$, the BPS string tensions are \begin{eqnarray*}
T_{\lambda_{1}} & = & a\pi m\left(n-1\right),\\
T_{\lambda_{n-1}} & = & T_{\lambda_{n}}=a\pi m\frac{n\left(n-1\right)}{4},\end{eqnarray*}
which results the string ratios\begin{equation}
\frac{T_{\lambda_{k}}}{T_{\lambda_{1}}}=\frac{C(\lambda_{k})}{C(\lambda_{1})}=\frac{n}{4}\,\,\,\,\,\,\textrm{for \,$k=n,\, n-1.$}\label{5.35a}\end{equation}
 Note that, for the $G=SU(n)$, for fixed $k$ and large $n$, the
ratio of the tensions coincide, for the Casimir and sine law scalings
with $T_{\lambda_{k}}=kT_{\lambda_{1}}$ for $n\rightarrow\infty$.
 In contrast, for $G=Spin(2n)$, the Casimir and sine law scalings
give different results in the large $n$ limit, as it can be seen
from Eqs. (\ref{5.25a}) and (\ref{5.35a}).

\subsubsection*{c) For $G=E_{6}$ }

For $E_{6}$, $h=12$ and \[
C(\lambda_{1})=C(\lambda_{5})=\frac{52}{3}.\]
Then, the BPS string tensions associated to $\lambda_{1}$ and $\lambda_{5}$
are\[
T_{\lambda_{1}}=T_{\lambda_{5}}=8a\pi m\]
 and \begin{equation}
\frac{T_{\lambda_{5}}}{T_{\lambda_{1}}}=\frac{C(\lambda_{5})}{C(\lambda_{1})}=1.\label{5.37}\end{equation}

\section{Conclusions}

In this work we analyzed the $Z_{N}$ string solutions in Yang-Mills-Higgs
theories with simple gauge groups $G$ spontaneously broken to their
center $C_{G}$. We studied two different vacuum solutions responsible
for the symmetry breaking $G\rightarrow U(1)^{r}\rightarrow C_{G}$,
for any $G$. We showed that for one vacuum, in the Coulomb phase,
the particles and monopoles of the theory with group $G$ have the
same masses as the particles and solitons of the corresponding Affine
Toda Field Theory, if the couplings of the two theories satisfy some
suitable relations, which may indicate a relation between these theories.
The same result holds for ${\cal {\cal N}}=4$ super Yang-Mills theories.
Then, we reviewed the construction of the asymptotic form of the $Z_{N}$
string solutions in the Higgs phase and the matching of the fluxes
of the $Z_{N}$ strings and monopoles for any $G$ and arbitrary vacuum
which produces the symmetry breaking (\ref{1}). We then showed that
for each of the two vacua, the ratios of the tensions of the minimal
BPS $Z_{N}$ strings (associated to the representation with smallest
tension for each topological sector) satisfy the Casimir scaling and
the sine law scaling for $G=SU(N)$ and we extended these scalings
for any simple gauge group $G$, analyzing in particular $G=Spin(2n)$
and $G=E_{6}$. For the sine law scaling, the tensions are proportional
to the components $x_{i}^{(1)}$ of the left Perron-Frobenius eigenvector
of $K_{ij}$ and the ratios of tensions are equal to the ratios of
soliton masses of the correponding affine Toda field theory, for any
gauge group $G$. For the Casimir scaling, we obtained that the ratios
of tensions were equal to the ratios of the second Casimir of the
fundamental representations associated to the different topological
sectors (\ref{5.40}). These results show that for the $Z_{N}$ strings,
these scalings are not {}``universal laws'', but they depend on
the vacuum which produce the symmetry breaking. From the dual superconductor
picture, this result may indicate that \textcolor{black}{tensions
of the QCD strings could be due to a non-Abelian monopole condensate
in one of these two vacua. }

The spectrum of QCD string tensions has been extensively studied in
recent years in lattice calculations\cite{lattice}\cite{latticeteper}.
In particular in \cite{latticeteper} it was observed that the QCD
string tensions lie between the Casimir and sine law scalings (a little
above the Casimir scaling). On the other hand, the Casimir scaling
(\ref{5.29d}) and the sine law scaling (\ref{5.19}) are lower bounds
for the non-BPS $Z_{N}$ string tensions and they hold exactly only
for the BPS $Z_{N}$ strings, which exist only on the boundary between
a type I and a type II superconductor. Therefore, the deviation from
the Casimir scaling observed in \cite{latticeteper} could be due
to fact that QCD strings would not be BPS. 

The properties analyzed so far for the $Z_{N}$ string solutions indicate
that they could be the magnetic analogous to QCD strings. We hope
that our results may be useful for lattice calculation for analyzing
the QCD strings with $G=SU(n)$ and other gauge groups. 

\vskip 0.2 in \noindent \textbf{}\textbf{\large Acknowledgments} \textbf{}

\noindent

I would like to thank Luigi Del Debbio for discussions.

\end{document}